# Quantum at a Music Festival: the Impact of an Exhibit about Quantum Science and Technologies on Festival Visitors

Vincent Koeman[1,2], Sanne Romp[1,3], Sanne J.W. Willems[4], Julia Cramer[1,2]

[1] *Science Communication & Society, Leiden University, Sylviusweg 72, 2333BE Leiden*
[2] *Leiden Institute of Physics, Leiden University, Niels Bohrweg 2, 2333CA Leiden*
[3] *Nationaal Expertisecentrum Wetenschap & Samenleving, Winthontlaan 2, 3526KV Utrecht*
[4] *Methodology & Statistics, Leiden University, Wassenaarseweg 52, 2333AK Leiden*

**Abstract:** Quantum technologies are seen as transformative, with a potential to revolutionize fields like drug discovery and machine learning. Public engagement is crucial to align these developments with societal needs and foster acceptance. This study measured the impact of an exhibit about quantum technologies at the 2024 Lowlands music festival ($n$ = 812). Pre- and post-surveys assessed changes in attitude, concern, interest and subjective knowledge. Results showed an increase in subjective knowledge but a decrease in interest, possibly due to reduced novelty or increased perceived difficulty. These findings underscore the effectiveness of exhibits as outreach tools in informal settings and highlight the critical role of maintaining novelty and emphasizing the relevance of quantum technologies in future outreach efforts. Additionally, we emphasize the importance of assessing outreach effectiveness to ensure that objectives are successfully achieved.

## Introduction

Science communication literature has advocated engaging the public with science and technology (Sturgis, 2014; Stilgoe et al., 2014). Such engagement can improve the development of informed policies and regulations for science and technology (Roberson, 2021; Weingart et al., 2021) and drive discussions around ethical, legal and social concerns (Weingart et al., 2021). In turn, these discussions are thought to help advance science and develop technologies in a responsible manner (Owen et al., 2012), mitigating risks and exploiting perceived benefits. Outreach can help with putting science and research central for innovation and building a sustainable future (Jensen & Gerber, 2020). Outreach encompasses all communication about scientific topics that is done outside of classrooms (Kim & Dopica, 2016; Sánchez-Mora, 2016). Participation in outreach may spark situational interest, a temporary form of engagement triggered by a specific experience, which may develop into long-lasting interest (Hidi & Renniger, 2006). In turn, this interest could cause long-term engagement (Hidi & Renniger, 2006).

Evidence-based outreach can contribute to increase the effectiveness and quality of outreach (Jensen & Gerber, 2020), for example through measuring the impact on the public's knowledge or attitude (Varner, 2014). Unfortunately, the practice of measuring outreach impacts remains uncommon (Volk & Schäfer, 2024; Volk, 2024). For instance, Volk (2024) found that out of 128 Swiss National Science Foundation (SNSF) funded science communication projects, only 8.6% used pre-/post-design studies to measure effects on knowledge and attitude. Specifically for emerging technologies, the effectiveness of outreach has not been studied often. However, studies on outreach for nanotechnologies (Duncan et al., 2010) and gene modification technologies (Rose et al., 2017) report positive changes in knowledge and attitude.

An important emergent technology currently under development for which the effect of outreach has not been studied to date is quantum technology. Quantum technologies are categorized in three major domains: quantum computing (Paudel et al., 2022), quantum networks (Wehner et al., 2018) and quantum sensors (Degen et al., 2017). These technologies are expected to have impact on our current society (Vermaas, 2017). Shorter time spans for drug discovery, quicker developments for machine learning, stronger cybersecurity and improved biomedical equipment are expected benefits of quantum technologies (De Wolf, 2017; Aslam et al., 2023). At the same time there are concerns about the monopolization surrounding these technologies (Seskir et al., 2023), unequal access to the technologies (Ten Holter et al., 2022) and misuse of quantum technologies (Busby et al., 2017). To better steer the development of these applications for society, it is important to engage the public at an early stage (Kop et al., 2023).

In this study, we focus on assessing the impact of a quantum intervention at a music festival. The following section provides a theoretical framework for this study, including relevant concepts and literature.

**Theoretical framework**

*1.1 Science outreach in a festival setting*

Informal settings, like music festivals, are strategic places to perform outreach due to the informal surroundings and free-choice learning (Sardo & Grand, 2016). Informal surroundings can positively affect people's attitude towards participating in outreach by surprising them (Sardo & Grand, 2016) and they can give people more confidence to actively participate during outreach activities (Bultitude & Sardo, 2012). Additionally, informal environments that are perceived as safe and stimulating can increase science festival visitors' level of interest in new scientific knowledge (Jensen & Buckley, 2014) leading to greater engagement (Harackiewicz et al., 2016). These informal environments often include an aspect of autonomy, leaving visitors free choice to engage with the science communication intervention. This autonomy has been found to positively influence their outreach experience (Falk et al., 2007; Sardo & Grand, 2016). This could lead to positive changes in interest, cognitive and emotional effects of the outreach activity. Moreover, free-choice learning and a leisure atmosphere have been shown to positively affect interest, learning and attitude towards the discussed topic (Reber et al., 2009, Storksdieck et al., 2005).

*1.2 Assessing outreach interventions*

Outreach efforts contribute to long-term public engagement through affecting four outcome variables: attitude, level of concern, interest and subjective knowledge (Volk, 2024). Attitude indicates the likelihood of visitors re-engaging with quantum technologies (Glasman & Albarracín, 2006). Visitors' level of concern may indicate awareness of the topic (Sjöberg, 2004). Interest shows visitors' motivation to explore the topic in the future (Hidi & Renninger, 2006). Finally, subjective knowledge plays a role in visitors' confidence and motivation to further engage with quantum (Wigfield & Eccles, 2000). By measuring the effect of our exhibit on these variables, we can gauge how the exhibit influences visitors' engagement with quantum technologies.

*Attitude*

A person's attitude towards a technology influences their likelihood of engaging with it in the future (Glasman & Albarracín, 2006). The Theory of Planned Behavior (Ajzen, 1991) suggests that attitudes shape behavioral intentions, which in turn influence actual behavior. In the case of quantum technologies, a more positive attitude may increase openness to learning, support for its development, and willingness to engage with related discussions.

Beyond likelihood of future engagement, a positive attitude towards a technology can improve public acceptance of that technology. New technologies often face skepticism or resistance (Samhan, 2018). These public perceptions are influenced by factors such as familiarity, media representation, and perceived societal impact (Kim & Kankanhalli, 2009). Studies suggest that when people are more engaged with novel technologies, they tend to form more positive attitudes and are more likely to accept new technologies (Roberson, 2021).

*Concern*

While attitude focuses on emotional responses towards new technologies, concern reflects potential reservations or apprehensions. Emerging technologies may elicit concerns among the general public, as seen in the case of human gene editing (Rose et al., 2017) and nanotechnologies (Gupta et al., 2015). These concerns play a role in influencing acceptance levels and shaping policy debates (Bearth & Siegrist, 2016). Often rooted in perceived risks, ethical considerations or broader societal implications, public concern can affect the degree of trust placed in scientific institutions (Wintterlin et al., 2022). Moreover, levels of concern can also be a measure for awareness about the topic (Sjöberg, 2004).

*Interest*

Interest plays an important role in fostering long-term engagement with science and technology. Defined as a tendency to engage with a specific topic now or in the future (Hidi & Renniger, 2006), interest is central in numerous models surrounding motivation, such as the expectancy-value theory (Wigfield & Eccles, 2000) and the self-determination theory (Ryan & Deci, 2000). These frameworks suggest that people are more interested in a topic when they perceive it as valuable, feel capable of understanding it and enjoy learning about the topic.

Interest can develop over time and with repeated exposure to a topic (Hidi & Renniger, 2006; Renniger & Hidi, 2020), making outreach efforts key opportunities to induce interest. Stand alone outreach efforts induce situational interest (Renniger & Hidi, 2020), which has been shown to enhance attention and engagement during an activity (Harackiewicz et al., 2016). With repeated and prolonged exposure to the topic, situational interest develops into individual interest (Renniger & Hidi, 2020). Individual interest is more ingrained in somebody's identity, making interest a good predictor of long-term engagement (Azevedo, 2013a; 2013b; McLaughlin et al., 2018).

*Subjective knowledge*

Higher levels of subjective knowledge increases likelihood of future engagement and influences attitude towards a topic. Subjective knowledge, or perceived understanding, refers to an individual's self-assessment of their knowledge about a topic. The connection between subjective knowledge and future engagement can be understood through motivational theories, for instance the expectancy-value theory (Eccles, 2009). This theory posits that motivation to engage with a topic is influenced by a perceived expectancy for success. In other words, people who feel more knowledgeable about a scientific topic may be more confident in engaging with it. Additionally, research suggests that subjective knowledge also influences attitude towards science (Fonseca et al., 2023).

*Science capital*

Science capital describes the extent to which participants already engage with and have access to science (Archer et al., 2015). Four domains are used to describe the aspects related to science capital: knowledge and literacy (e.g. understanding of sciences and the scientific method), science-related dispositions and beliefs (e.g. perceptions of usefulness of science), behaviours and tendencies (e.g. how people already engage with science in their daily lives), and social circles (e.g. family or friends that work or study in science) (Archer et al., 2015). Through science capital, more detail can be acquired about the audience of outreach events (Archer et al., 2015). Specifically, it can provide insights into why some people might be able to utilize new scientific information more than others (Kaakinen et al., 2025). Thus, science capital might influence changes in attitude, interest and subjective knowledge brought about by outreach efforts.

*1.3 The present research*

In this study, we assess the impact of outreach on quantum science and technologies in the form of a pop-up exhibit at a music festival on the engagement of its visitors.

To answer this question, we pose the following subquestions:

- What is the effect of an exhibit about quantum science and technologies on visitors' attitude, concern, interest and subjective knowledge regarding quantum science and technologies?
- How does the visitors' level of science capital influence the impact of an exhibit about quantum science and technologies?

This study was preregistered (https://osf.io/9rcyn) and approved by the Ethics Review Committee of the Faculty of Science at Leiden University (Number: 2024 - 013). We openly share all study materials, anonymized data, analysis scripts and supplemental tables (https://osf.io/mpf8n/files/osfstorage).

**Materials and Methods**

To assess the impact of the quantum pop-up exhibit at a music festival on the engagement of its visitors, we measured visitors' changes in attitude, concern, interest and subjective knowledge through a pre- and post-survey study at Lowlands Science during the annual Lowlands Music festival (MOJO, 2025a) in 2024. In this section, the Lowlands Music Festival and the exhibit are introduced. Afterwards, we discuss the participants, outline the procedure, and specify the measures used in our survey. We close of this section by describing our analysis.

*Lowlands Music Festival*

Lowlands is organized yearly in late August in Biddinghuizen, The Netherlands. This study was performed during the 32$^{nd}$ edition in 2024, from the 16$^{th}$ until the 18$^{th}$ of August. This festival is visited by more than 60,000 visitors annually (Docter-Loeb, 2024). Since 2015, part of the festival grounds has been reserved for Lowlands Science, where researchers can apply for performing research projects (MOJO, 2025b). During the 2024 edition, 10 studies have been conducted at Lowlands Science (Docter-Loeb, 2024). Visitors need to buy tickets for the Lowlands festival, but can then freely enter the Lowlands Science terrain between 12.00h and 20.00h, walk around on these grounds and decide for themselves if and when they participated in one of the studies.

*Quantum: The pop-up exhibit*

"Quantum: The pop-up exhibit" is a bilingual exhibit (in English and Dutch; Institute for Quantum Computing, n.d.). It was designed in 2016 by the Institute For Quantum Computing of Waterloo, Canada (in English and French). The goal of the exhibit is to communicate the impact, importance and opportunities of quantum science and technologies to various audiences in Canada. Discussed quantum technologies include quantum computers, quantum sensors and quantum networks, and explained quantum science concepts include superposition and the wave-particle duality. The exhibit comprises 13 panels which use text, pictures, two videos and four interactives to explain quantum technology concepts and how these concepts could shape our future (Institute for Quantum Computing, n.d.). Pictures of the exhibit can be found in the Supplementary Material (Appendix 1).

Seven of the panels have an explanatory role; introducing the exhibit, describing two rules of quantum science, the essence of quantum, quantum in our daily lives, quantum technology and qubits, quantum cryptography and quantum sensors. The other six panels can be interacted with in various manners. One panel includes a digital simulation of the double slit experiment to explain the wave-particle duality. Another features a rotatable polarizer through which a standing or lying cat could be seen to illustrate superposition using Schrödinger's cat. One of two video panels introduces the impact of quantum technologies, the other explains how quantum computing works. A hands-on puzzle allows visitors to explore how classical computer bits function. Lastly, a flip-board panel contrasts the computing power of quantum computers with that of classical computers.

*Participants*

Participants in this study were Lowlands Science visitors aged 16 or older that voluntarily passed by the exhibit and participated without being solicited or persuaded. Participation in the study was not a prerequisite to enter the exhibit. Prior knowledge about quantum technologies was not a prerequisite to partake in this study. The participants were included in this study when they signed an informed consent form before the start of the research, in which they agreed on using their anonymous data in this research. As participants were visitors to a music festival, there is a possibility that they were under the influence of drugs or alcohol (Geuens et al., 2022). Participants were excluded from the study if they were visibly intoxicated.

The study participants (N = 812) had a mean age of about 31 ($SD_{Age}$ = 9.5), 53.3% were male (N = 407), 45.3% were female (N = 346) and 0.9% reported as a third gender of non-binary (N = 9). Most of the participants were highly educated; 76,9% (N = 758) noted they have a degree from a university of applied sciences or a university. This group made up 36,4% of the Netherlands between 15 and 75 years of age (CBS, 2024). Therefore, the participants are not representative of the entire Dutch population. Further demographics can be found in Supplementary Material (Table S1).

*Procedure*

People at Lowlands Science were encouraged to participate in this research by engaging them in small conversations about quantum technologies at the Lowlands Science terrain, without explaining concepts that are addressed in the exhibit. At the entrance of the exhibition, visitors received a short spoken explanation of the research, filled out an informed consent form and the pre-survey on paper, and received a random code to connect the pre-survey to the post-survey. Hereafter, visitors could enter the pop-up exhibit. There was no mandatory walking route and visitors could spend as long as they wanted in the exhibit, typically around 20 minutes. After finishing their visit, visitors filled out the post-survey.

*Measures*

The pre-survey, in Dutch and English, consisted of three demographic questions and five sets of 5-point Likert scale items measuring science capital, interest, attitude, concern and subjective knowledge. If not stated otherwise, the answer to the Likert scale items all ranged from 1 (completely disagree) to 5 (completely agree).

The first set of 5-point Likert scale items consisted of four statements measuring the participants' science capital (e.g. "I am generally aware of new scientific discoveries and developments"), adopted from Peeters et al. (2022). The next three items measured the participants' level of interest (e.g. "I want to know more about how quantum technologies work"), based on questions used in the studies of Joubert et al. (2020) and Shulman & Sweitzer (2018). To measure visitors' attitude towards quantum technologies four items were used (e.g. "Quantum technologies are exciting"), adapted from the Attitude Towards Science Inventory by Tai et al. (2022). Concern was measured using three items modified from Gardner & Troelstrup (2015; e.g. "I am concerned about quantum technologies in general"). At the end of the survey, four questions were posed to gauge the participants' subjective knowledge about the topics which were discussed in the exhibit (e.g. "How confident are you in your understanding of superposition?"). The answers to these items ranged from 1 (completely unconfident) to 5 (completely confident).

The post-survey contained the same questions as the pre-survey, but they were focused on the participants' view after their visit to the exhibit (e.g. "After visiting the exhibit, I think quantum technologies are exciting"). First, the set of items about attitude was asked followed by the set about concern and interest about quantum technologies (e.g. "After visiting the exhibit, I want to know more about how quantum technologies work"). Next, participants were asked to answer the four statements about their confidence in their knowledge on several aspects of quantum science and technologies (e.g. "After visiting the exhibit, how confident are you in your understanding of superposition?").

The complete pre- and post-surveys can be found in the Supplement Material (Appendix 2). Some items were reverse-coded such that high scores on the survey correspond to positive or high levels of science capital, interest, attitude, concern and subjective knowledge. If the Cronbach's alpha scores showed internal consistency of the items, the means of the participants' responses in the pre- and post-test were used as composite scores for each scale.

*Statistical Analysis*

The paper surveys were manually digitized using Excel (version 2016). To ensure accuracy in data entry, a second coder reliability check was performed on 10% of the data, yielding a 99% agreement rate. All data analysis and visualization were conducted using R (version 4.3.2).

To test whether attitude, concern, interest and subjective knowledge changed after seeing the exhibit, we used paired, two-sided t-tests to compare the mean scores of the outcome measures in the pre- and post-test. Given the large sample size (812 participants), the t-tests are assumed to be robust against non-normality. However, we did not assume that the variances were equal and used the Welch approximation to the degrees of freedom.

To check whether participants' science capital was associated with the effects of the exhibit, their Pearson correlation with the changes in attitude, concern, interest and subjective knowledge were calculated. Additionally, to find unexpected correlations we performed exploratory analyses by calculating the Pearson correlation between science capital and the measures prior to the exhibit just as between the measures prior to the exhibit and the difference between the pre- and post-survey results.

## Results

*Science capital*

As illustrated in Figure 1, participants demonstrated relatively high levels of science capital, reporting high scores for awareness of scientific developments (*M* = 3.4, *SD* = 1.1), interest in the scientific process (*M* = 4.1, *SD* = 0.9), engaging in science-related activities during their free time (*M* = 3.7, *SD* = 1.0) and discussing science with friends or colleagues (*M* = 3.5, *SD* = 1.2).

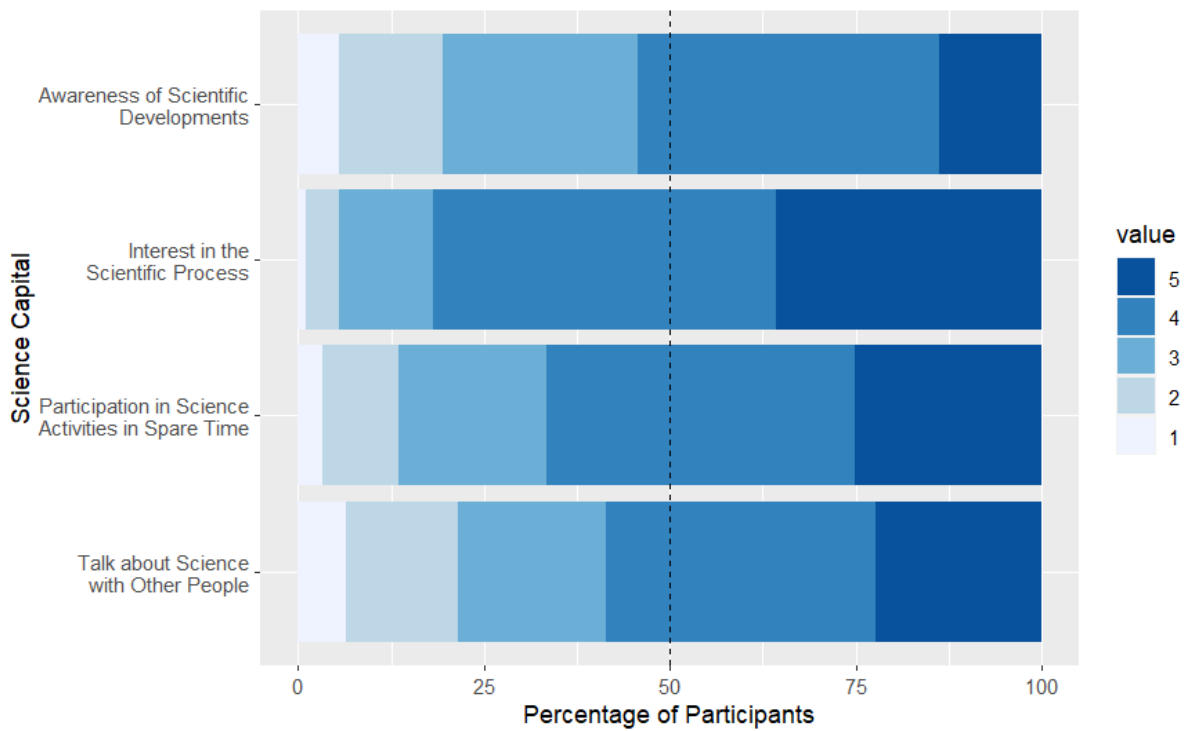

**Figure 1.** Stacked bar graph showing the participants' answers to the items related to science capital. The graph illustrates four items: (1) habit to talk about science, (2) participate in science activities in their spare time, (3) their level of interest in the scientific process and its results, and (4) their awareness of scientific developments.

*Change in attitude, concern, interest and subjective knowledge*

Overall, the internal consistency of the pre- and post-survey were adequate. The Cronbach's alphas can be found in the Supplementary Material (Table S2). Only in the set measuring the participants' level of concern weak consistency was found ($\alpha_{pre}$ = .44, $\alpha_{post}$ = .60). By omitting the item "I have no concerns about quantum technologies because the benefits are likely greater than the risks" the internal consistency improved ($\alpha_{pre}$ = .69, $\alpha_{post}$ = .77). Therefore, this item was omitted and the two remaining items were used to calculate the mean scores.

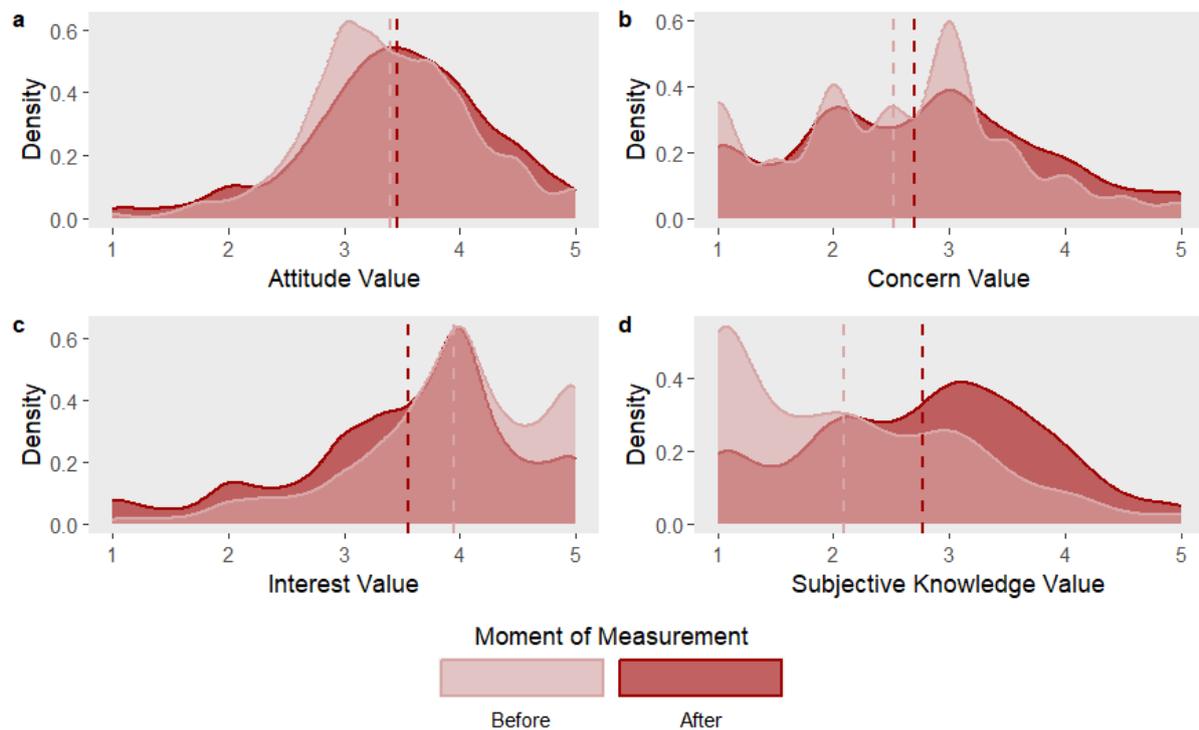

**Figure 2.** Density plots of the distribution of the composite scores of the outcome measures (a) attitude, (b) concern, (c) interest and (d) subjective knowledge. The vertical dashed lines represent the mean values before and after visiting the exhibit.

A distribution of the scores on the pre- and post-survey for each measure is shown in Figure 2, and the results of the *t*-test comparing these scores are presented in Table 1. We observe a significant, but very small (*d* = 0.08) positive shift in the participants' attitude towards quantum technologies between the pre-test and post-test. The participants had a neutral level of concern prior to visiting the exhibit. After visiting the exhibit, their level of concern increased. The increase was significant, but the effect size was small (*d* = 0.18). The exhibit seemed to have a medium effect (*d* = 0.44) on visitors' interest in quantum technologies, with a significant decrease after visiting the exhibit as compared to before their visit to the exhibit. A significant and large increase (*d* = 0.80) was found in the participants' subjective knowledge about quantum science and technologies after they visited the exhibit.

**Table 1.** Results of the paired t-tests.

| Variable | $M_{pre}$ | $SD_{pre}$ | $M_{post}$ | $SD_{post}$ | t | Degrees of Freedom | *p* | Cohen's *d* |
|---|---|---|---|---|---|---|---|---|
| **Attitude** | 3.4 | 0.7 | 3.5 | 0.8 | 2.18 | 790 | 0.030 | 0.08 |
| **Concern** | 2.5 | 1.0 | 2.7 | 1.1 | 5.08 | 793 | <0.001 | 0.18 |
| **Interest** | 3.9 | 0.8 | 3.5 | 1.0 | -12.34 | 794 | <0.001 | 0.44 |
| **Subjective Knowledge** | 2.1 | 1.0 | 2.8 | 1.0 | 21.61 | 728 | <0.001 | 0.80 |

*Association between science capital and attitude, concern, interest and subjective knowledge*

Pearson correlation coefficients were calculated to find out plausible linear correlations between science capital and the changes in attitude, concern, interest and subjective knowledge. Weak correlations were found with attitude ($r(785) = -.15$, $p < .001$), interest ($r(790) = -.13$, $p < .001$) and subjective knowledge ($r(723) = -.09$, $p = .015$). No significant correlation was found between science capital and the change in concern ($r(788) = -.06$, $p = .089$). Inspection of the scatterplots depicting the relationships between science capital and the changes in interest, concern, attitude and subjective knowledge revealed no evidence of nonlinear relations. These figures can be found in the Supplementary Material (Appendix 3).

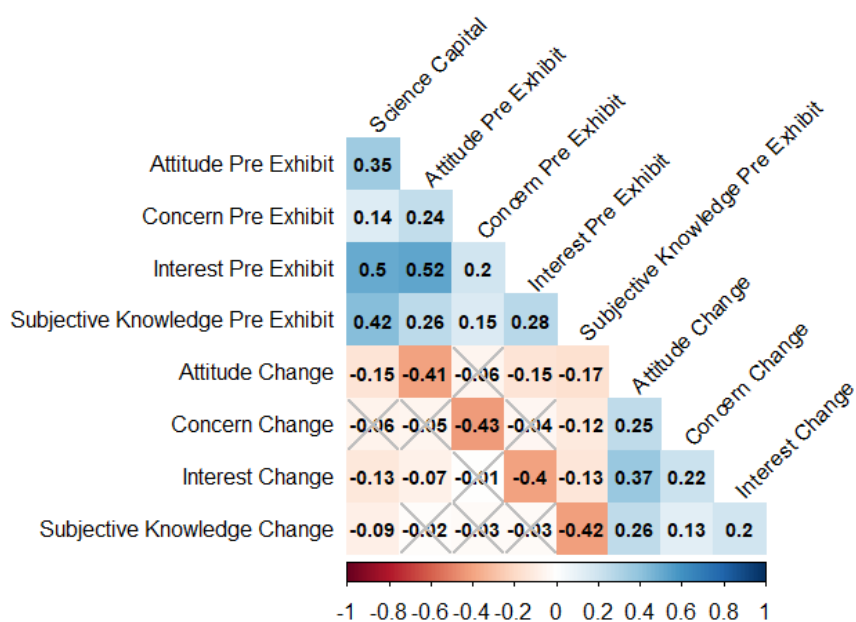

**Figure 3.** Pearson correlation coefficients matrix. Correlation coefficients that are crossed-out are insignificant ($p > 0.05$).

*Exploratory analyses*

We further explored correlations between the pre-measures and the measured changes, a complete overview can be found in Figure 3. A strong, positive correlation was found between the participants' level of interest before visiting the exhibit and the participants' attitude before visiting the exhibit ($r(795) = 0.52$, $p < 0.001$). A moderate, positive correlation was found between participants' change in interest and their change in attitude ($r(781) = 0.37$, $p < 0.001$).

**Discussion**

This study examined the impact of a pop-up exhibit on quantum science and technology on the engagement of visitors of the exhibit at the Lowlands music festival. By comparing 812 participants' survey responses before and after engaging with the exhibit, we assessed effects on attitude, concern, interest and subjective knowledge and studied their relation to science capital. These findings offer valuable insights into how outreach efforts can influence public understanding of quantum science and technologies. Below we discuss the significance of the results and their implications for the exhibit. We finish this paper by discussing some limitations of this study and providing a future outlook.

*Negligible effect on attitude*

The data indicated a positive, but very small, effect on the visitors' attitude towards quantum science and technologies after a visit to the exhibit. This means the exhibit had little immediate effect on visitors' attitudes towards quantum science and technologies. This small effect is consistent with other findings (Sripaoraya et al., 2022; Gall et al., 2020; Darienzo et al., 2024).

It is unsurprising that we found a relatively small change in attitude, as a visitor needs to experience incongruencies between their prior beliefs and new information for attitude changes to occur (Stone & Taylor, 2021). Since visitors are not expected to perceive many of these incongruencies in a brief exhibit, it is unlikely that major attitudinal changes will be measured. Additionally, visitors reported a low level of knowledge about quantum technologies prior to visiting the exhibit. Therefore, visitors might have had few prior beliefs around the topic and, hence, experienced few incongruencies around the topic (Stone & Taylor, 2021). In the future, as quantum technologies near implementation and public awareness increases, the pop-up exhibit may lead to greater shifts in attitudes, since people are likely to be more informed about these emerging technologies.

Moreover, the interactivity of the exhibit could have influenced the results of our study. Comparing the exhibit with a presentation activity by Sripaoraya et al. (2022) where participants were involved in presenting science, we find that the interactivity of our quantum exhibit is lower. This might explain the smaller change in attitude in this study. Interactivity in science communication has been shown to have a positive influence on the audiences' attitude towards the discussed topic (Sundar & Kim, 2005).

*Minor increase in concern*

Visiting the exhibit increased visitors' concerns about quantum technologies slightly. This is in line with findings of other studies (Rose et al., 2017; Dijkstra & Critchley, 2016). Since concern can be conceptualized as a form of attitude, the same rhetoric can be followed as used for the small change in attitude.

As in the example of Rose et al. (2017), both the attitude and perceived concern of the visitors are increased due to their visit to the exhibit. These results may reflect the exhibit's ability to communicate risks and benefits of quantum technologies. Showing the risks and benefits of new technologies increases trust in the provided information (Satterfield et al., 2013), which could lead to long term engagement with quantum technologies (Satterfield et al., 2013). Given that one of the exhibit's central aims was to raise awareness of quantum technologies' societal implications, the simultaneous increase in concern and positive attitude among visitors suggests that this goal was effectively met.

*Moderate decrease in interest*

After visiting the exhibit, the participants reported a moderately decreased interest in learning about quantum technologies. Similar studies found contrasting results. For instance, at CERN, a study was done to find out the effect of the "S'Cool Lab", a science outreach lab focusing on hands-on experiments in particle physics (Woithe et al., 2022). After a visit, high school students were more interested as compared to before the exhibit. Additionally, in the study of Meinsma et al. (2024), the authors found that participants were more interested in quantum technologies after they were presented with a short text about quantum technologies in which the underlying quantum phenomena were explained.

A possible explanation for the decline in interest in our study can be described by a decrease in perceived novelty and a high perceived difficulty of quantum science and technologies. A perceived novelty towards a topic increases interest and motivation to learn more (Sung et al., 2016; González-Cutre, 2016). Quantum technologies are a recent development, therefore, we can consider that there is a high perceived novelty around quantum technologies, as reflected in our data by a low level of subjective knowledge prior to visiting the exhibit. This high perceived novelty could increase the participants' interest in these topics prior to the exhibit. However, the exhibit might demystify quantum technologies, decreasing the perceived novelty of the topic and, in turn, decreasing the interest in quantum technologies.

The decrease in interest could also be due to an increase in the perceived difficulty of the topic or the texts used in the exhibit. Fulmer and Tulis (2013) tested the effect of perceived difficulty of a text among middle school students. They found that perceived difficulty and interest in a topic were negatively correlated; when people perceive a text as more difficult their interest in the topic is decreased. Similarly, Nuutila et al. (2021) found that higher perceived difficulty can negatively affect situational interest. It has been shown that interest in a novel topic is influenced by the visitors' perceived ability to understand the topic (Noordewier & Van Dijk, 2016; Silvia, 2005).

*Substantial increase in subjective knowledge*

Before visiting the exhibit, visitors reported having a low level of subjective knowledge ($M_{pre}$ = 2.1, $SD_{pre}$ = 1.0). After the exhibit, visitors reported a substantial increase in their subjective knowledge ($M_{post}$ = 2.8, $SD_{post}$ = 1.0). Other studies showed similar results after their outreach events. One study showed that participants perceived learning during a Pint of Science event focused on infection, microbes and doing research (Adhikari et al., 2019). Additionally, Sripaoraya et al. (2022) found that people had more confidence in their ability to learn science and Rose et al. (2017) found an increase of 0.62 points, on a 5-point Likert scale, regarding their participants' perceived knowledge after visiting a panel on human gene editing. Taken together, the exhibit's impact on subjective knowledge mirrors findings from other outreach activities, indicating that outreach efforts are effective means of increasing perceived knowledge.

*Limitations*

One key limitation of this study lies in the context of the music festival setting. Visitors voluntarily chose to engage with the exhibit, likely indicating a pre-existing interest in science, which may limit the generalizability of the findings to broader populations. Additionally, the festive environment, characterized by alcohol and drug use (Geuens et al., 2022) may have influenced participants' cognitive processing and emotional responses (Magrys & Olmstead, 2014). Although the research team screened for obvious signs of intoxication, it is likely that some intoxicated individuals still participated. The environment of a music festival could have positively biased participants' responses due to heightened mood (Lee, 2014), which we did not measure or control for.

Methodologically, some limitations should be noted. Moreover, the study did not track the duration of engagement with the exhibit, which prevents an analysis of how time spent might correlate with observed changes. The short-term, single-visit and self-reported nature of the research method also limits insights into the durability of any observed effects. Prior studies suggest that outreach impacts may fade over time, especially when interventions are brief and lack interactivity (Fletcher et al., 2021; Bailey et al., 2020). Finally, some Cronbach's alpha values fell below the acceptable threshold of 0.70.

*Future research*

Future research could build on our findings in several directions. The contrast between the observed decrease in interest and Meinsma et al.'s (2024) reported increase following a brief explanatory text about quantum science and technologies suggests that the amount and type of explanation may influence interest differently. Future studies could systematically manipulate explanation depth across formats to explore this interaction. Another promising avenue is investigating the role of perceived novelty, as quantum science's often "mystical" framing (Meinsma et al., 2023) may shape public interest. Qualitative methods, such as visitor interviews, could also shed light on which exhibit components drive engagement with quantum science and technologies, helping to reveal the cognitive and emotional mechanisms that underlie effective science communication.

Additionally, further research should consider how context affects visitor engagement and learning about quantum technologies. Our setting may have introduced confounding factors such as noise, crowding, or a recreational mindset, all of which could influence experiences. Comparing outcomes across diverse environments like museums, schools, or science centres would clarify how setting shapes attention, interest, and knowledge retention.

*Practical implications*

Our findings suggest that an exhibit at a music festival may serve as an effective means to enhance visitors' perceived knowledge of quantum science and technologies. However, to strengthen the observed effects, several adjustments could be considered in the design and content of such exhibits.

Quantum technologies could be made more relatable by explicitly linking quantum technologies to everyday life, current issues and personal experiences to foster greater interest among a broader audience (Giamellaro, 2017). Making these connections more tangible may help visitors see the relevance of quantum science in everyday life. Previous research supports these directions. For example, Noordewier and Van Dijk (2016) found that comparing new technologies to familiar ones can enhance interest and coping potential by increasing perceived familiarity (Rindova & Petkova, 2007).

While the current exhibit presents a broad range of quantum technologies and underlying principles, this breadth may have introduced too much novelty at once, potentially overwhelming visitors. Research by Cors et al. (2018) suggests that excessive novelty can hinder engagement, while an appropriate balance between novelty and familiarity enhances learning. Focusing more narrowly on one or two core concepts could reduce cognitive overload, increase perceived knowledge of specific topics, and promote deeper understanding.

In conclusion, this study offers an insight in the extent to which outreach about quantum science and technology impacts festival visitors. We utilised pre-post surveys to observe mixed outcomes including an increase in subjective knowledge and a decrease in interest. These findings highlight the nuanced challenges of engaging the public with emerging technologies, such as quantum technologies. While our results are limited by the context of a festival environment, they highlight the importance of interactivity and perceived relevance of the exhibit. Based on these findings we recommend for future outreach to narrow the scope of their outreach and focus on making quantum technologies more relatable. Future research should explore how the explanation depth and the context affect the outreach's outcome.

**Funding**

The Dutch National Growth Fund (NGF), as part of the Quantum Delta NL programme, finances the data collection.

**Acknowledgements**

This work was supported by the Dutch National Growth Fund (NGF), as part of the Quantum Delta NL programme, and the Strategic Communication and Marketing directorate at the Leiden University.

We are grateful to the Lowlands organisation (NWA, BKB) for the opportunity to conduct this study at the Lowlands festival. Our sincere thanks also go to the Institute for Quantum Computing, particularly John Donohue, for generously allowing us to borrow the exhibit.

We appreciate the dedication and support of our research team during the festival, including Liselotte Rambonnet, Nienke Beets, Michelle Willebrands, Julie Schoorl, Anna Heerdink, Muhammed Karami and Brenda Roovers. Additionally, thank the staff of *New Scientist* for their efforts in promoting our study and outreach before and during the festival.

We acknowledge Aletta Meinsma and Francien Bossema for their insights, which have contributed to this article. Lastly, we would like to extend our gratitude to Ivo van Vulpen and Sense Jan van der Molen for their support throughout the project.